\documentclass[twocolumn,showpacs,preprintnumbers,amsmath,amssymb,prb]{revtex4}

\usepackage{epsfig}

\newcommand{\cumu}[2]{\langle\!\langle #1 \rangle\!\rangle_{#2}}
\newcommand{\expct}[2]{\left\langle #1 \right\rangle_{#2}}
\newcommand{\T}{\mathcal{T}}
\newcommand{\D}{\mathbb{D}}

\newcommand{\SSigma}{\bf{\Sigma}}

\begin{document}

\title{Full counting statistics of spin transfer through the Kondo dot}
\author{T.~L.~Schmidt,$^{1,3}$ A.~O.~Gogolin,$^1$ and A.~Komnik$^{2,3,4}$}
\affiliation{${}^1$Department of Mathematics, 180 Queen's Gate,
London SW7 2AZ, United Kingdom \\
${}^2$Service de Physique Th\'eorique, CEA Saclay, F–-91191
Gif-sur-Yvette, France \\
${}^3$Physikalisches Institut, Albert--Ludwigs--Universit\"at
Freiburg, D--79104 Freiburg, Germany\\
${}^4$Institut f\"ur Theoretische Physik, Universit\"at
Heidelberg, D--69120 Heidelberg, Germany}
\date{\today}

\begin{abstract}
We calculate the spin current distribution function for a Kondo dot
in two different regimes. In the exactly solvable Toulouse limit the
linear response, zero temperature statistics of the spin transfer is
trinomial, such that all the odd moments vanish and the even moments
follow a binomial distribution. On the contrary, the corresponding
spin-resolved distribution turns out to be binomial. The combined
spin and charge statistics is also determined. In particular, we
find that in the case of a finite magnetic field or an asymmetric
junction the spin and charge measurements become statistically
dependent. Furthermore, we analyzed the spin counting statistics of
a generic Kondo dot at and around the strong-coupling fixed point
(the unitary limit). Comparing these results with the Toulouse limit
calculation we determine which features of the latter are generic
and which ones are artifacts of the spin symmetry breaking.
\end{abstract}


\maketitle

\section{Introduction}

It has been known for some time that the current noise spectra of
mesoscopic systems can reveal information about transport processes
which is not accessible via measurement of the average current
alone. Indeed, it has been shown by Schottky \cite{schottky18} that
the characteristics of charge carriers can be investigated by
measuring noise spectra. The noise is the second moment of the
distribution function which gives the probability that a certain
amount of charge is transported within a certain time interval $\T$.
Recent developments in experimental mesoscopic physics have
facilitated the measurement of even higher order moments of this
distribution.\cite{reulet03} Therefore, it is desirable to calculate
the complete statistics, which is often referred to as the full
counting statistics (FCS). The FCS has been found for the charge
transport in non-interacting systems by Levitov and Lesovik
\cite{levitov93} and, currently, interaction effects are under
intensive investigation.

Recent years have witnessed a soaring interest in the field of
spintronics,\cite{awschalom02} where the principal objective is to
gain a level of control over spin transport (which comprises
creation of current and its measurement) comparable to that over
charge transport. Indeed, the usage of the electron spin degree of
freedom offers many enticing opportunities which are currently being
investigated. Therefore, in order to gain a thorough understanding
of spin-related processes taking place in mesoscopic systems, one
has to understand the properties of spin current to the same level
of accuracy as the charge current.

A method of measuring spin currents and some intriguing consequences
of consecutive measurements of different spin components have been
presented by Di Lorenzo and Nazarov.\cite{lorenzo04} Although some
general statements about spin transfer FCS have been made there, to
the best of our knowledge, there has so far been no derivation of
the spin transfer FCS for a concrete system. The aim of this article
is to fill this gap. While the charge FCS was shown to be binomial,
at least in the universal linear response, zero temperature
limit,\cite{gogolin06} it is not immediately clear what the nature
of the spin statistics should be: the natural guess that it is the
inverse binomial distribution\cite{mandel95} (due to bosonic nature
of spin excitations) turns out to be wrong. The purpose of this
article is to investigate the spin transfer FCS for the transport
through a Kondo dot, where non-zero spin currents can be relatively
easily generated by applying a finite magnetic field and a bias
voltage. Furthermore, we go one step further and calculate the
combined FCS of charge and spin transfer, which will allow us to
examine cross-correlations of these two currents and to obtain
spin-resolved charge transfer statistics.

We shall perform this analysis on two distinct systems which
represent two aspects of the Kondo physics. First, we will
investigate a Kondo dot at the Toulouse point, where for a certain
choice of coupling constants, the system can be diagonalized exactly
by means of bosonization and subsequent refermionization. Second, we
will go to the strong-coupling regime where the dot spin is
hybridized with the leads spins to form a singlet state. Virtual
excitations of this singlet state can be treated perturbatively and
will be shown to produce a finite spin current noise. This article
will close with a comparison of these two regimes.

\section{Hamiltonian}

The system under investigation consists of two non-interacting leads
and an interacting dot level. The left/right leads are being held at
chemical potentials $\mu_{L/R}$, respectively, where $V = (\mu_L -
\mu_R)/e$ is the applied voltage. The position of the bare dot level
and the electron-electron repulsion are such that the dot is always
occupied by exactly one electron. The result is the two-channel
Kondo Hamiltonian,\cite{schiller98} $H = H_0 + H_K + H_B + Y_0$,
where (in units of $e=\hbar=v_F=1$),
\begin{eqnarray}\label{H_Kondo}
  H_0 & = & i \sum_{\alpha = LR} \sum_{\sigma = \uparrow\downarrow}
            \int dx \psi_{\alpha \sigma}^\dag(x) \partial_x \psi_{\alpha\sigma}(x)
            \, , \nonumber \\
  H_K & = & \sum_{\alpha \beta = LR} \sum_{\nu=x,y,z} J^{\alpha \beta}_\nu
  s^\nu_{\alpha\beta} \tau^\nu  \, , \nonumber \\
  H_B & = & - \mu_B g h \tau^z =: -\Delta \tau^z \, , \nonumber \\
  Y_0 & = & \frac{V}{2} \sum_{\sigma=\uparrow\downarrow} \int dx
  \left[ \psi_{L\sigma}^\dag \psi_{L\sigma} - \psi_{R\sigma}^\dag
  \psi_{R\sigma} \right] \, .
\end{eqnarray}
In this equation, $\tau^{x,y,z}$ are proportional to the Pauli
matrices describing the impurity spin, $J^{\alpha\beta}_\nu$ for
$\alpha, \beta = L,R$ and $\nu = x,y,z$ are coupling constants for
the possible transport processes, $h$ is the magnetic field applied
to the dot level. Furthermore,
\begin{equation}
    \vec{s}_{\alpha\beta} = \frac{1}{2} \sum_{\sigma,\sigma'}
    \psi_{\alpha\sigma}^\dag(0) \vec{\tau}_{\sigma\sigma'}
    \psi_{\beta\sigma'}(0)\, ,
\end{equation}
for $\alpha,\beta = L,R$ are the components of the electron spin
density at the impurity site.

The aim of FCS is the calculation of the cumulant generating
function (CGF) $\chi(\mu,\lambda)$ for the probability distribution
function $P(s,q)$ to transfer $q$ units of charge and $s$ units of
spin during the waiting time $\T$. In our case, as we are interested
in the combined statistics of spin transfer and charge transfer, we
define $\chi(\mu,\lambda) = \sum_{s, q} e^{i s \mu} e^{i q \lambda}
P(s,q)$. Partial derivatives of this function then allow us to
calculate arbitrary correlation functions of charge and spin
transport and cross-correlations between both.

In order to calculate spin and charge current in this system, we
proceed as usually in tunneling setups and define charge and spin
flavor number operators
\begin{eqnarray}
  N_c & = & \frac{1}{2} \sum_{\alpha=LR} \sum_{\sigma=\uparrow\downarrow}
  \alpha \int dx \psi^\dag_{\alpha\sigma}(x) \psi_{\alpha\sigma}(x)\, , \\
  N_s & = & \frac{1}{2} \sum_{\alpha=LR} \sum_{\sigma=\uparrow\downarrow}
  \alpha \sigma \int dx \psi^\dag_{\alpha\sigma}(x) \psi_{\alpha\sigma}(x)\, .
\end{eqnarray}
The time derivatives of these operators are proportional to the
charge and spin currents, respectively. Hence, the corresponding
charge and spin currents can be calculated by means of the
Heisenberg equation $I_{c,s} = i[H,N_{c,s}]$. The general approach
to calculate higher order cumulants of these operators involves
including fictitious measurement devices for both charge and spin
current in the Hamiltonian.\cite{nazarov03} This leads to additional
terms $H_m = \mu I_s + \lambda I_c$. As in our case the particle
number operators $N_c$ and $N_s$ commute, both terms can be removed
by a unitary transformation. The counting fields $\mu$ and $\lambda$
then appear as exponentials in those components of the Hamiltonian
which transfer spin or charge in either direction:\cite{levitov04}
from the left lead to the right, $T^{c,s}_{L}$, and backwards,
$T^{c,s}_{R}$. As in this case the underlying processes for spin and
charge transport will be different, the tunneling Hamiltonian
separates into spin and charge parts and becomes [see
Eq.~(\ref{T_lambda_mu1})]
\begin{eqnarray}
    T^{\mu\lambda} & = & e^{i\mu(t)} T^s_L + e^{-i\mu(t)} T^s_R
    \nonumber \\
    & + & e^{i\lambda(t)} T^c_L + e^{-i\lambda(t)} T^c_R \, ,
\end{eqnarray}
where $\mu(t)$ and $\lambda(t)$ are explicitly time-dependent
functions defined on the Keldysh contour $C$. The counting fields
are only ``switched on'' during the measurement time $\T$\!, such
that we can define them as $\mu(t_\pm) = \pm \mu \theta(t) \theta(\T
- t)$ and analogously for $\lambda(t_\pm)$. The CGF is then given by
the expectation value\cite{levitov04}
\begin{equation}\label{chi1}
    \chi(\mu,\lambda) = \expct{T_C \ e^{-i\int_C ds
    T^{\mu\lambda}(s)}}{},
\end{equation}
where $T_C$ denotes the contour ordering operator. In order to
calculate this function, we employ the generalized Green's function
formalism that has previously been used successfully to calculate
the FCS of other interacting systems: \cite{komnik05,gogolin06} We
assume $\lambda(s)$ and $\mu(s)$ to be arbitrary functions defined
on $C$ that have slowly varying Keldysh components $\lambda_\pm(t)$
and $\mu_\pm(t)$, respectively. Assuming $\T$ to be large, we can
neglect processes which are due to the switching of the counting
fields, and we can introduce the adiabatic potential $U(\mu_\pm,
\lambda_\pm)$ which entails
\begin{equation}\label{chi2}
    \chi[\mu_\pm(t), \lambda_\pm(t)] = e^{-i\int_0^\T dt U[\mu_\pm(t),
    \lambda_\pm(t)]}\, ,
\end{equation}
from which the statistics can be recovered using the equation $\ln
\chi(\mu, \lambda) = -i \T U(\mu, \lambda)$ as $U(\mu, \lambda)$
will turn out to be time-independent. By comparison of (\ref{chi1})
and (\ref{chi2}), one notices that the adiabatic potential can be
calculated due to the following relation (which is a consequence of
the Feynman-Hellmann theorem)
\begin{equation}\label{dUdlambda}
    \frac{\partial U}{\partial \lambda_-} = \expct{\frac{\partial
    T^{\mu\lambda}}{\partial \lambda_-}}{\mu\lambda}
\end{equation}
or, equivalently, via the partial derivative with respect to
$\mu_-$. The expectation value with subscript $\mu\lambda$ is
defined as expectation value in the interaction picture with respect
to the tunneling Hamiltonian $T^{\mu\lambda}$,
ie.\cite{komnik05,gogolin06}
\begin{equation}
    \expct{A(t)}{\mu\lambda} = \frac{1}{\chi(\mu_\pm, \lambda_\pm)}
    \expct{T_C\left\{ A(t) e^{-i\int_C ds T^{\mu\lambda}(s)}
    \right\}}{}.
\end{equation}
The emerging expression is slightly more complicated than the usual
Hamiltonian formalism. However, the great advantage of using the
derivatives of $U$ is the emergence of Green's functions (the time
dependence of which is dominated by $T^{\mu\lambda}$) for which a
Dyson equation can be obtained. Then, the adiabatic potential
$U(\mu, \lambda)$ can be recovered by integrating (\ref{dUdlambda})
(or, equivalently, the derivative with respect to $\mu_-$, the
result is the same) which, in turn, leads immediately to the CGF
$\chi(\mu, \lambda)$.

\section{Dyson equation at the Toulouse~point}
For a general constellation of system parameters,
$\chi(\mu,\lambda)$ cannot be calculated exactly. The
perturbation theory in the exchange couplings is
well known to diverge logarithmically and so is not
very useful in calculating the FCS.
However, it has been realized by Toulouse that the
Kondo Hamiltonian can be reduced to an easily integrable shape for a
special parameter set. \cite{toulouse69} In our notation this
so-called Toulouse point is attained for $J^{\alpha\beta}_x =
J^{\alpha\beta}_y =: J^{\alpha\beta}_\perp, J^{LR}_\perp =
J^{RL}_\perp$ and $J^{LR}_z = J^{RL}_z = 0$.\cite{schiller98} The
only allowed processes are then: (i) spin-flip tunneling which
transports one unit of charge but does not contribute to the spin
current, and (ii) spin-exchange between one lead and the dot which
only transports one unit of spin with no associated charge
transport. Furnishing both processes with the respective counting
fields $\mu(t)$ and $\lambda(t)$ the tunneling operator becomes
\begin{eqnarray}\label{T_lambda_mu1}
  T^{\mu\lambda} & = &
  \frac{J^{LL}_\perp}{2} \Big[ \psi_{L\uparrow}^\dag \psi_{L\downarrow} e^{-i\mu/2} \tau^- + \psi_{L\downarrow}^\dag \psi_{L\uparrow}  e^{i\mu/2} \tau^+\Big]
  \nonumber \\
  & + &
  \frac{J^{RR}_\perp}{2} \Big[ \psi_{R\uparrow}^\dag \psi_{R\downarrow} e^{i\mu/2} \tau^- + \psi_{R\downarrow}^\dag \psi_{R\uparrow}  e^{-i\mu/2} \tau^+\Big] \nonumber  \\
  & + &
  \frac{J^{LR}_\perp}{2} \Big[ \psi_{L\uparrow}^\dag \psi_{R\downarrow} e^{-i\lambda/2} \tau^- + \psi_{L\downarrow}^\dag \psi_{R\uparrow} e^{-i\lambda/2} \tau^+ \nonumber \\
  & + &
  \psi_{R\uparrow}^\dag \psi_{L\downarrow} e^{i\lambda/2} \tau^- + \psi_{R\downarrow}^\dag \psi_{L\uparrow} e^{i\lambda/2} \tau^+
  \Big].
\end{eqnarray}
In the next step, we proceed along the lines of Schiller and
Hershfield:\cite{schiller98} First, we bosonize the Hamiltonian by
introducing four bosonic fields describing total charge (c), charge
flavor (f), total spin (s) and spin flavor (sf). Then, we perform an
Emery-Kivelson\cite{emery92} rotation and refermionize the
Hamiltonian. After going to the Toulouse point $J^{LL}_z = J^{RR}_z
=: J_z := 2\pi$, the total spin field decouples and the resulting
Hamiltonian is quadratic in the three remaining fermionic fields.
The unperturbed part reads
\begin{eqnarray}
    H_0 & = & i \sum_\nu \int dx \psi^\dag_\nu(x) \partial_x \psi_\nu(x)\, , \nonumber \\
    H_B & = & \Delta(d^\dag d - 1/2)\, , \nonumber \\
    Y_0 & = & V \int dx \psi^\dag_f(x) \psi_f(x)\, ,
\end{eqnarray}
where $d = i \tau^+$ is the fermion operator describing the dot spin
and $\nu \in \{c, f, sf\}$. The tunneling part becomes
\begin{eqnarray}\label{T_lambda_mu2}
  T^{\mu\lambda} & = &
  -\frac{J^{LL}_\perp}{2\sqrt{2\pi a_0}} \left[ e^{-i\mu/2} d^\dag \psi^\dag_{sf} - e^{i\mu/2} d
  \psi_{sf}\right] \nonumber \\
& - &
  \frac{J^{RR}_\perp}{2\sqrt{2\pi a_0}} \left[ e^{i\mu/2} d^\dag \psi_{sf} - e^{-i\mu/2} d \psi^\dag_{sf}\right] \nonumber  \\
& - &
  \frac{J^{LR}_\perp}{2\sqrt{2\pi a_0}} \Big[ e^{-i\lambda/2} d^\dag \psi^\dag_f +
  e^{-i\lambda/2} d \psi^\dag_f  \nonumber \\
& - & e^{i\lambda/2} d^\dag \psi_f - e^{i\lambda/2} d\psi_f \Big]\,
,
\end{eqnarray}
where $a_0$ is the lattice constant of the underlying lattice model
which emerges in the bosonization process. Note that
$T^{\mu\lambda}$ only contains $\psi_f$ and $\psi_{sf}$ which means
that $\psi_{c}$ is completely decoupled and can safely be discarded.
Physically, this is due to the fact that the total charge in
conserved in the system. Next, we split the relevant fields into
hermitian components (Majorana fermions) according to $\psi_{sf} =
(\xi + i \eta)/\sqrt{2}$, $\psi_f = (\zeta + i \epsilon)/\sqrt{2}$,
$d = (a + i b)/\sqrt{2}$, and obtain the Hamiltonian
\begin{equation}
    H = H_0 + i V \int dx \, \zeta(x) \epsilon(x) -i\Delta a b +
    T^{\mu\lambda}\, ,
\end{equation}
where the tunneling operator is given by
\begin{eqnarray}\label{T_lambda_mu3}
    T^{\mu\lambda} & = & iJ_- \sin(\mu/2) a\xi + iJ_+ \cos(\mu/2) b\xi \nonumber \\
    &+ &iJ_- \cos(\mu/2) a\eta
  - iJ_+ \sin(\mu/2) b\eta \nonumber \\
  & + & i J_\perp \sin(\lambda/2) a \zeta + i J_\perp \cos(\lambda/2) a
  \epsilon\, .
\end{eqnarray}
In this equation, we have introduced $J_\pm = (J^{LL}_\perp \pm
J^{RR}_\perp)/(2\sqrt{2\pi a_0})$ and $J_\perp =
J^{LR}_\perp/(\sqrt{2\pi a_0})$. According to (\ref{dUdlambda}), the
next step is to calculate the expectation value of the partial
derivatives of $T^{\mu\lambda}$ which can easily be calculated from
(\ref{T_lambda_mu3}). If we calculate the derivative with respect
to, say, $\lambda_-$,
\begin{equation}\label{derivTlambda}
    \expct{\frac{\partial T^{\mu\lambda}}{\partial \lambda_-}}{\mu\lambda}
    = -\frac{J_\perp}{2} \big[ \cos(\lambda_-/2) G^{--}_{a \zeta} -
    \sin(\lambda_-/2) G^{--}_{a \epsilon}\big]\, ,
\end{equation}
we obtain an expression which contains the Green's functions
\begin{eqnarray}\label{Ga}
    G^{--}_{a\zeta}(t,t') & = & -i\expct{T a(t) \zeta(t')}{\mu\lambda}
    \label{Gaz}  \\
    G^{--}_{a\epsilon}(t,t') & = & -i\expct{T a(t) \epsilon(t')}{\mu\lambda}
    \label{Gae}
\end{eqnarray}
where the time arguments $t, t'$ are located on the ``$-$''-branch
of the Keldysh contour and, consequently, are time-ordered.
Expanding (\ref{Gaz}) to first order in $J_\perp$, one can express
it in terms of the exact dot Green's function $D_{aa}(t,t') =
-i\expct{T_C a(t) a(t')}{\mu\lambda}$. Assuming the counting fields
$\mu_\pm(t)$ and $\lambda_\pm(t)$ to be time-independent on the
respective branches of the Keldysh contour, we have
\begin{eqnarray}
    G^{--}_{a\zeta}(\omega)
& = & iJ_\perp \Big[
    s_- D^{--}_{aa} G^{(0)--}_{\zeta\zeta} +
    c_- D^{--}_{aa} G^{(0)--}_{\epsilon\zeta} \nonumber \\
& - &
    s_+ D^{-+}_{aa} G^{(0)+-}_{\zeta\zeta} -
    c_+ D^{-+}_{aa} G^{(0)+-}_{\epsilon\zeta}\Big] \, ,
\end{eqnarray}
where $s_\pm = \sin(\lambda_\pm/2)$ and $c_\pm =
\cos(\lambda_\pm/2)$. An analogous equation can be derived for
$G_{a\epsilon}$. The exact dot Green's function can be derived by
means of a Dyson equation which in Fourier space reads $\D =
\D^{(0)} + \D \SSigma \D^{(0)}$. The matrix of dot Green's function
is defined by
\begin{equation}\label{def_D_matrix}
    \D = \left(
           \begin{array}{cccc}
             D^{--}_{aa} & D^{-+}_{aa} & D^{--}_{ab} & D^{-+}_{ab} \\
             D^{+-}_{aa} & D^{++}_{aa} & D^{+-}_{ab} & D^{++}_{ab} \\
             D^{--}_{ba} & D^{-+}_{ba} & D^{--}_{bb} & D^{-+}_{bb} \\
             D^{+-}_{ba} & D^{++}_{ba} & D^{+-}_{bb} & D^{++}_{bb} \\
           \end{array}
         \right)
\end{equation}
and the self-energy $\SSigma$ is defined as a block matrix coupling
the Majorana components $a$ and $b$ of the dot operator,
\begin{equation}
    \SSigma = \left(
           \begin{array}{cc}
             M_{aa} & M_{ab} \\
             M_{ba} & M_{bb} \\
           \end{array}
         \right)\, .
\end{equation}
It turns out that this matrix only depends on $\lambda := (\lambda_-
- \lambda_+)/2$ and $\mu := (\mu_- - \mu_+)/2$. We therefore
introduce the shorthand notation $s_{\mu,\lambda} =
\sin(\mu,\lambda)$ and $c_{\mu,\lambda} = \cos(\mu,\lambda)$. The
non-diagonal elements are then given by
\begin{equation}
  M_{ab} =  J_- J_+ \left(
                         \begin{array}{cc}
                           0                                & -s_\mu G^{(0)-+}_{\xi\xi} \\
                           s_\mu G^{(0)+-}_{\xi\xi} & 0 \\
                         \end{array}
                       \right)
\end{equation}
and $M_{ba} = -M_{ab}$. Next, we have for the $M_{bb}$ element
\begin{equation}
    M_{bb}
    =  J_+^2
    \left(
      \begin{array}{cc}
        G^{(0)--}_{\xi\xi}                  & -c_\mu G^{(0)-+}_{\xi\xi} \\
        -c_\mu G^{(0)+-}_{\xi\xi}   &  G^{(0)++}_{\xi\xi} \\
      \end{array}
    \right)\, .  \label{Maa}
\end{equation}

\begin{widetext}
These three matrices do not involve the charge counting field
$\lambda$ which is a consequence of the fact that the
$\lambda$-dependent part of the tunneling operator
(\ref{T_lambda_mu3}) only couples to the Majorana fields $a$. The
only $\lambda$-dependent contribution in the Dyson equation arises
from the $M_{aa}$ component which is given by
\begin{eqnarray}
    M_{aa} & = & J_-^2
    \left(
      \begin{array}{cc}
         G^{(0)--}_{\xi\xi}    & -c_\mu G^{(0)-+}_{\xi\xi} \\
        -c_\mu G^{(0)+-}_{\xi\xi}   &   G^{(0)++}_{\xi\xi} \\
      \end{array}
    \right)+
J^2_\perp
    \left(
      \begin{array}{cc}
          G^{(0)--}_{\zeta\zeta}   & - c_\lambda G^{(0)-+}_{\zeta\zeta} + s_\lambda G^{(0)-+}_{\epsilon\zeta}\\
        - c_\lambda G^{(0)+-}_{\zeta\zeta} -s_\lambda G^{(0)+-}_{\epsilon\zeta}   &   G^{(0)++}_{\zeta\zeta} \\
      \end{array}
    \right) \, .
\end{eqnarray}
\end{widetext}
As the zero order lead and dot Green's functions are well
known,\cite{komnik05} solving the Dyson equation amounts to
inverting a $4\times 4$-matrix. The ensuing calculation is lengthy
but straightforward. The result is an expression for the exact dot
Green's function, which can be used to evaluate
(\ref{derivTlambda}). The equation can then be integrated to obtain
the adiabatic potential $U(\mu,\lambda)$ which in turn determines
the CGF.

\section{Cumulant generating function}

\begin{figure}[t]
    \centering
    \includegraphics{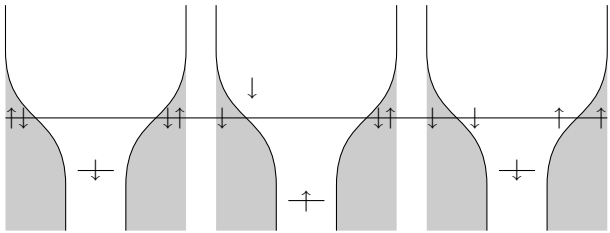}
    \caption{Schematic representation of the process $T_2$.}
    \label{T2diag}
\end{figure}

Finally, one obtains the CGF which describes the combined statistics
of charge and spin transport and which is the main result of this
article,
\begin{eqnarray}\label{lnchi_mu_lambda}
  \ln \chi(\mu, \lambda) & = & \T \int_0^\infty
    \frac{d\omega}{2\pi} \ln\bigg\{ 1 +  \frac{1}{D(\omega)}
  \Big\{ \kappa (e^{i 2 \mu} + e^{-i 2 \mu}-2)\nonumber
  \\
& + &
  \sum\limits_{p=\pm}\Big[ \alpha_p (e^{i 2 p \lambda} - 1) +
  \nu_p (e^{i \mu} - e^{- i \mu}) e^{i p
  \lambda} \nonumber \\
& + &
  \beta_p \left( e^{i (\mu + p \lambda)} + e^{- i (\mu - p
  \lambda)}- 2\right) \nonumber \\
& + & \rho_p \left(e^{i 2 (\mu + p \lambda)} + e^{- i 2 (\mu - p
\lambda)}
  - 2\right) \Big]
 \Big\} \bigg\}\, ,
\end{eqnarray}
where $\Gamma_i = J_i^2/2$ and
\begin{eqnarray}
    D(\omega) & = & \left[ \omega^2 - \Delta^2 - \Gamma_+(\Gamma_\perp +
    \Gamma_-)\right]^2 \nonumber \\
& + &
    \omega^2\left(\Gamma_+ + \Gamma_- +
    \Gamma_\perp\right)^2.
\end{eqnarray}
Defining $n_F(\omega), n_R(\omega)$ and $n_L(\omega)$ to be the
Fermi functions with chemical potentials $0$, $-V/2$ and $V/2$,
respectively, the `transmission coefficients' are given by
\begin{eqnarray}\label{lnchi_mu_lambda_coeff}
  \alpha_+& = &  n_L (1 - n_R) \left[ 1 + 2n_F(n_F-1)\right]
  \Gamma_+^2 \Gamma_\perp^2 \nonumber \\
  &+& \omega^2 \, n_L (1 - n_R) \Gamma_\perp^2 \nonumber \, , \\
  \beta_+  & = &  \left[n_F(1-n_R) + n_L(1-n_F)\right] \Gamma_+
  \Gamma_\perp (\Gamma_- \Gamma_+ + \Delta^2) \nonumber \\
  &+& \omega^2 \,  \left[n_F(1-n_R) + n_L(1-n_F)\right]
  \Gamma_- \Gamma_\perp \nonumber \, , \\
  \kappa  & = &  \left[ n_R(1-n_L) + n_L(1-n_R) - 1\right]
  n_F (n_F - 1) \Gamma_+^2 \Gamma_\perp^2 \nonumber \\
  &+& \omega^2   n_F (1 - n_F) \left( \Gamma_- - \Gamma_+\right)^2 \nonumber
  \, , \\
  \rho_+  & = &  n_L( 1 - n_R) n_F (1 - n_F) \Gamma_+^2 \Gamma_\perp^2 \nonumber \, ,
  \\
  \nu_+   & = &  2\omega \left[ n_F(n_L - n_R + 1) - n_L \right]
  \sqrt{\Gamma_- \Gamma_+} \Gamma_\perp
  \Delta \, ,
\end{eqnarray}
where all the quantities with opposite subscripts are given by the
same expressions with interchanged $R$ and $L$ indices. From
(\ref{lnchi_mu_lambda}), one can easily calculate all correlation
functions of charge and spin currents. In particular, by setting
$\mu=0$, one immediately recovers the known charge transfer
statistics for this system.\cite{komnik05} Before returning to the
complete statistics (\ref{lnchi_mu_lambda}), we shall investigate in
more detail the spin transfer statistics. By setting $\lambda = 0$,
we obtain the CGF for the spin transfer,
\begin{eqnarray}\label{lnchi_T}
    \ln \chi(\mu) & = & \T \int_0^\infty \frac{d\omega}{2\pi} \ln \bigg\{
    1 +  T_{1a}  (n_F - n) (e^{i\mu} - e^{-i\mu}) \nonumber \\
    & + & T_{1s} \left[ n(1-n_F) + n_F(1-n) \right]  (e^{i\mu} +
e^{-i\mu} -
    2) \nonumber \\
    & + & T_2 n_F(1-n_F)(e^{2i\mu} + e^{-2i\mu} - 2) \bigg\}
\end{eqnarray}
where $n = (n_L + n_R)/2$ and the `transmission coefficients' are
given by
\begin{eqnarray}
  T_{1a}(\omega) & = & \frac{4}{D(\omega)}
  \left[\sqrt{\Gamma_- \Gamma_+} \Gamma_\perp
  \Delta\omega \right], \nonumber \\
  T_{1s}(\omega) & = & \frac{2}{D(\omega)}\left[\Gamma_+ \Gamma_\perp \left(\Gamma_- \Gamma_+ + \Delta^2\right) +
  \Gamma_- \Gamma_\perp\omega^2 \right],\nonumber \\
  T_2(\omega) & = & \frac{1}{D(\omega)}
  \left[ \Gamma_+^2 \Gamma_\perp^2 + (\Gamma_- -
  \Gamma_+)^2\omega^2 \right].
\end{eqnarray}
One can see that two kinds of processes contribute to the
statistics. The coefficients $T_{1s}$ and $T_{1a}$ represent
symmetric and anti-symmetric tunneling events in which one unit of
spin is transferred. Therefore, $T_{1a}$ controls the odd cumulants,
in particular the average spin current whereas $T_{1s}$ only
influences even cumulants like the noise. Odd cumulants obviously
vanish for a symmetric system ($\Gamma_-=0$) or zero magnetic field
($\Delta=0$). On the other hand, $T_2$ involves voltage-independent
processes which transfer two units of spin. Those are coupled
spin-flip processes which do not involve electron tunneling and can
only occur at finite temperatures, see Fig.~\ref{T2diag}.

\section{Special regimes}

\subsection{Zero temperature case}
Next, we shall address some special regimes and begin with the zero
temperature case. In the limit of low energy we obtain the following
statistics,
\begin{equation}\label{chi_T0V0}
    \chi(\mu) =\Big[
    (1 - T_{1s}) + \frac{1}{2} T_{1s} \left(e^{i\mu} +  e^{-i\mu}\right)
    \Big]^N.
\end{equation}
This is basically a symmetric trinomial distribution. Restoring SI
units, the exponent can be written as $N = eV\T/h$, in which case
$N$ can be interpreted as the number of `attempts',\cite{levitov96}
i.e. the number of electrons impinging on the tunneling contact in
time $\T$. Comparing it with the corresponding distribution function
for the charge transfer,\cite{komnik05} one concludes that
(\ref{chi_T0V0}) is the same distribution with odd moments removed.
Hence, the lowest order average spin current is quadratic in $V$,
precisely as found by Schiller and Hershfield.\cite{schiller98} The
noise in the zero temperature limit, however, is linear in $V$,
\begin{eqnarray}\label{noise_T0_Esmall}
  \cumu{\delta s^2}{}
  & = & \frac{\T V}{\pi} \frac{\Gamma_+ \Gamma_\perp \left(\Gamma_- \Gamma_+ +
  \Delta^2\right)}{\left[ \Delta^2 + \Gamma_+(\Gamma_\perp +
    \Gamma_-)\right]^2} \nonumber \\
  & = & \frac{\T V}{\pi} T_0 \left[1- T_0\right] \, ,
\end{eqnarray}
where $\delta s = \int_0^\T dt I_s(t)$ in the spin transferred
during the measurement time $\T$. $T_0$ is an effective transmission
coefficient defined by $T_0 := \frac{1}{2} T_{1s}(0) + T_2(0)$ which
also emerges in the finite temperature case as we shall see soon.
Equation (\ref{noise_T0_Esmall}) is a manifestation of the general
formula for the shot noise power of a tunneling junction with
transmission coefficient $T_0$. The same leading order dependence on
the applied voltage has already been obtained
perturbatively.\cite{kindermann05} Moreover, in this limit, the same
formula has already been obtained for the charge
current.\cite{komnik05}

\begin{figure}[t]
    \centering
    \includegraphics{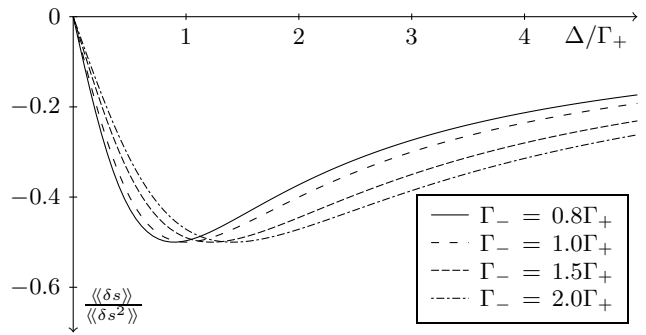}
    \caption{Current-noise ratio in units of $V/\Gamma_+$ for the
    spin current as a function of applied magnetic field}
    \label{fig_ratio}
\end{figure}

Due to the different behavior of current and noise as a function of
the applied voltage, one obtains a voltage-dependent current-noise
ratio
\begin{equation}
    \frac{\cumu{\delta s}{}}{\cumu{\delta s^2}{}}
=
    -\frac{V\sqrt{\Gamma_- \Gamma_+} \Delta}
    {\Gamma_+ (\Gamma_- \Gamma_+ + \Delta^2)} \, .
\end{equation}
One notices that for constant magnetic field the ratio becomes more
favorable for large bias voltages. The dependence on the magnetic
field for constant bias voltage is plotted in Fig.~\ref{fig_ratio}.
Notably, the function has a minimum at $\Delta =
\sqrt{\Gamma_- \Gamma_+}$. Towards large magnetic fields, the ratio
vanishes.

We can go one step further and calculate the third cumulant in the
same limit, which is quadratic in $V$. Normalized with respect to
the average spin current,
\begin{eqnarray}
  \frac{\cumu{\delta s^3}{}}{\cumu{\delta s}{}}
  & = &
  6\frac{\Gamma_+ \Gamma_\perp \left(\Gamma_- \Gamma_+ +
  \Delta^2\right)}{\left[\Delta^2 +
\Gamma_+(\Gamma_\perp + \Gamma_-)\right]^2} -1 \, ,
\end{eqnarray}
which can be interpreted as a generalized Fano factor.

\subsection{Equilibrium case}
The other easily accessible regime is the equilibrium case $V=0$.
Here, we trivially expect all odd cumulants to vanish. Indeed,
setting $n_L = n_R$ in (\ref{lnchi_T}), we notice that the
anti-symmetric part $T_{1a}$ vanishes and the CGF becomes
completely symmetric. The lowest order cumulant is therefore the
second one for which one obtains exactly the same Johnson-Nyquist
result as for the charge current fluctuations
\begin{equation}\label{noise_JN}
   \cumu{\delta s^2}{} = 4 G_0 \T k_B T T_{0},
\end{equation}
where $G_0 = e^2/h$ is the conductance quantum. By direct
calculation one verifies that in spite of finite magnetic field
the particle currents carrying different spins are indeed
completely uncorrelated.

\section{Cross-correlations}
\begin{figure}[t]
    \centering
    \includegraphics{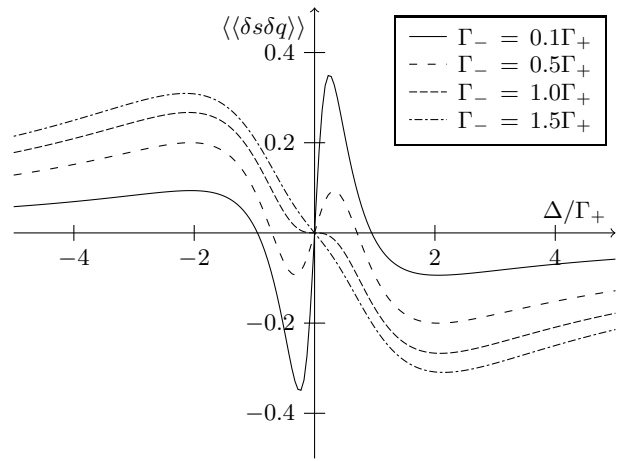}
    \caption{Normalized spin-charge current correlation
    in units of $V/\Gamma_+$ as a function of magnetic field
    for different values of $\Gamma_-$}
    \label{figure_correlation}
\end{figure}

Finally, we shall make use of the availability of the combined FCS
to obtain the correlation between charge current and spin current.
The quantity $\cumu{\delta s \delta q}{}$ can easily be obtained
from (\ref{lnchi_mu_lambda}) by double partial derivation with
respect to $\mu$ and $\lambda$. Again, for zero temperature and low
energies, we obtain the normalized correlation
\begin{eqnarray}
    \frac{\cumu{\delta s \delta q}{}}{\sqrt{\cumu{\delta s^2}{}  \cumu{\delta q^2}{}}}
    & = & \frac{V \sqrt{\Gamma_-\Gamma_+} \Gamma_\perp \Delta}
    {\left[ \Delta^2 + \Gamma_+(\Gamma_- + \Gamma_\perp)\right]^2}
    \nonumber \\
    & \times &
    \frac{\Gamma_+^2\Gamma_\perp^2 - \left(\Gamma_- \Gamma_+ + \Delta^2\right)^2}
    {\Gamma_+\Gamma_\perp  \left(\Gamma_- \Gamma_+ +
    \Delta^2\right)}\, .
\end{eqnarray}
This correlation function is plotted in
Fig.~\ref{figure_correlation}. It can become negative and, in
particular, changes its sign whenever the sign of $V$ or $\Delta$ is
switched. This was expected as the substitution $\Delta \rightarrow
-\Delta$ inverts the spin current while leaving the charge current
unchanged whereas substituting $V \rightarrow -V$ acts analogously
on the charge current. Moreover, we notice that there is a choice of
parameters, $\Gamma_+ \Gamma_\perp = \Delta^2 + \Gamma_- \Gamma_+$,
where this correlation becomes non-trivially zero. The correlation
of spin current and charge current comes about as a result of the
coupling of both currents to the dot level. Setting $\lambda = \mu =
0$ in (\ref{T_lambda_mu3}), we notice that such a coupling can be
removed by setting $\Gamma_- = \Delta = 0$. In this case, it can
indeed be shown that $\chi(\mu,\lambda) = \chi_s(\mu)
\chi_q(\lambda)$.

At $T=0$, in linear response the combined statistics takes an
appealing form,
\begin{equation}\label{comb00}
\chi(\lambda,\mu)=\left[1+T_1(\cos\mu \, e^{i\lambda}-1)+T_2
(e^{2i\lambda}-1)\right]^N \;,\nonumber
\end{equation}
where we recover the same transmission coefficients that had already
been found in the case of charge transport\cite{komnik05} and which
are responsible for electron and electron pair tunneling,
\begin{eqnarray}
    T_1 & = & 2 \Gamma_+ \Gamma_\perp (\Gamma_- \Gamma_+ +
    \Delta^2)/D(0), \nonumber \\
    T_2 & = & \Gamma_+^2 \Gamma_\perp^2/D(0).
\end{eqnarray}
We can recover the familiar binomial distribution if we go over to a
CGF for transfer of electrons with preselected spin orientation. To
that end we introduce a set of different counting fields with
self-evident notation $\lambda_{\uparrow, \downarrow} = \lambda \pm
\mu$. Then in analogy to the reduction presented in
[\onlinecite{komnik05,gogolin06}] we obtain
\begin{eqnarray}    \nonumber
 \chi(\lambda_{\uparrow, \downarrow}) = \left\{ \left[ 1 + T_e
 (e^{i \lambda_\uparrow} - 1) \right] \left[ 1 + T_e (e^{i
 \lambda_\downarrow} - 1) \right] \right\}^N \, ,
\end{eqnarray}
where the effective transmission coefficient is defined as $T_e =
\sqrt{T_2}$. This trivialization, however, takes place in the linear
response regime only. Nevertheless, this is a remarkable result
since even in the spin-resolved case the statistics of electron
transfer turns out to be determined by a single parameter $T_e$.

\section{Strong-coupling fixed point}
\label{unitarylimit} At temperatures below the Kondo temperature
$T_K$, the dot spin hybridizes with the lead spins and forms a
singlet state. At this so-called strong-coupling fixed point, the
system can be described as a pure resonant level system. The free
Hamiltonian can be written down in terms of non-interacting $s$- and
$p$-wave electrons, $c_{p\sigma}$ and $a_{p\sigma}$
respectively,\cite{hewson93}
\begin{eqnarray}\label{H0}
    H_0 & = &
    \sum_{p\sigma} \epsilon_{p\sigma} \left( c^\dag_{p\sigma}
    c_{p\sigma} + a^\dag_{p\sigma} a_{p\sigma} \right) \nonumber \\
    & + & \frac{V}{2} \sum_{p\sigma} \left( c^\dag_{p\sigma}
    a_{p\sigma} + a^\dag_{p\sigma} c_{p\sigma} \right)\, ,
\end{eqnarray}
where $\epsilon_{p\sigma} = \epsilon_p + \sigma h$ includes the
magnetic field $h$ and $V$ is the voltage difference between the two
leads. The lowest order excitations away from the strong-coupling
fixed point are given by a scattering term and an interaction term,
\begin{eqnarray}
        H_{sc} & = & \frac{\alpha}{2 \pi \nu T_K} \sum_{p p' \sigma}
        (\epsilon_p + \epsilon_p') c^\dag_{p \sigma} c_{p' \sigma}\,
        ,
\\
        H_{int} & = & \frac{\phi}{\pi \nu^2 T_K} \, ( c^\dag_{\uparrow}
        c_{\uparrow} - n_0)( c^\dag_{\downarrow} c_{\downarrow} -
        n_0)\, ,
\end{eqnarray}
where $n_0 = \expct{c^\dag_\sigma c_\sigma}{0}$ stands for the
particle density at $V=0$ and $h=0$. Moreover, $\nu$ denotes the
density of states and $\alpha$ and $\phi$ are dimensionless
amplitudes. In the actual Kondo model the Fermi-liquid relation
$\alpha = \phi = 1$ holds.\cite{nozieres74} Standard perturbation
theory in $\alpha$ and $\phi$ shows that no spin current is to be
expected. Therefore, we will focus on higher moments which are
conveniently accessible by means of full counting statistics.

For this purpose we include the counting fields for charge and spin
current. The most general approach involves the insertion of a
spin-resolved charge counting field $\lambda_\sigma$. Then, the
approach of [\onlinecite{gogolin06_2,schmidt07}] can easily be
generalized to
\begin{eqnarray}\label{acrot}
    c_\sigma(\lambda_\sigma) & = & \cos(\lambda_\sigma/4) c_\sigma - i \sin(\lambda_\sigma/4) a_\sigma\, ,\nonumber \\
    a_\sigma(\lambda_\sigma) & = & -i \sin(\lambda_\sigma/4) c_\sigma + \cos(\lambda_\sigma/4)
    a_\sigma\, .
\end{eqnarray}
The choice of this rotation is motivated by the fact that
re-inserting the first order expansion in $\lambda_\sigma$ into the
$H_{sc}$ and $H_{int}$ exactly gives the charge and spin currents
produced by these terms. We shall denote by
\begin{equation}
    H_\lambda = H_{sc}[c(\lambda)] + H_{int}[c(\lambda)]
\end{equation}
the perturbation Hamiltonian with rotated $c_\sigma$ and $a_\sigma$
operators. One can easily see that this rotation leaves $H_0$
invariant.

The term $H_0$ does not couple the spins and therefore, the
unperturbed statistics is a trivial combination of the statistics
for a spin-polarized system. Indeed, it is easy to show that
\begin{equation}
    \chi_0(\lambda_\sigma) = e^{i N (\lambda_\uparrow +
    \lambda_\downarrow)}\, ,
\end{equation}
where $N = (\T V)/(2\pi) = e V \T/h$ is the number of electrons of a
certain spin impinging on the impurity during the measurement time
$\T$. In order to address the lowest-order corrections in $\alpha$
and $\phi$, we start from the general expression
\begin{equation}
    \chi(\lambda_\sigma) = \chi_0(\lambda_\sigma) \expct{T_C
    \exp\left[ -i \int_C dt H_\lambda(t) \right]}{}\, .
\end{equation}
A linked cluster expansion to the lowest contributing order in
$H_\lambda$ leads to
\begin{eqnarray}
    \ln \chi(\lambda_\sigma)
& = & i N (\lambda_\uparrow +
    \lambda_\downarrow) \nonumber \\
& - & \frac{1}{2} \int_C dt_1 dt_2 \expct{T_C
    H_\lambda(t_1) H_\lambda(t_2) }{} \, .
\end{eqnarray}
This expression contains terms proportional to $\alpha^2$, $\phi^2$
and $\alpha \phi$. As the scattering Hamiltonian $H_{sc}$ does not
couple up- and down-spins, the spin-resolved statistics is again a
trivial combination of the two orientations,
\begin{widetext}
\begin{eqnarray}
    \delta_\alpha \ln \chi  =
    \frac{\alpha^2 \T V}{48 \pi T_K^2}
    \frac{V^2 + 4(\pi T)^2 + 12 h^2}{\sinh(V/2T)}
    \sum_\sigma \left[ (e^{-i\lambda_\sigma} - 1) e^{V/2T} +
    (e^{i\lambda_\sigma} - 1) e^{-V/2T} \right]\, .
\end{eqnarray}
By changing the basis of the counting fields and introducing
counting fields for spin and charge current, $\mu =
(\lambda_\uparrow - \lambda_\downarrow)/2$ and $\lambda =
(\lambda_\uparrow + \lambda_\downarrow)/2$, respectively, one can
readily perform the sum over $\sigma$ and one obtains
\begin{eqnarray}
    \delta_\alpha \ln \chi  =
    \frac{\alpha^2 \T V}{24 \pi T_K^2}
    \frac{V^2 + 4(\pi T)^2 + 12 h^2}{\sinh(V/2T)}
    \left\{ [\cos(\mu) e^{-i\lambda} - 1] e^{V/2T} +
    [\cos(\mu) e^{i\lambda} - 1] e^{-V/2T} \right\}\, .
\end{eqnarray}
Being even in $\mu$, this expression will not produce any spin
current but it will contribute to the spin current noise, albeit in
a trivial way. More interesting effects can be expected from the
interaction terms. Investigating the terms proportional to $\phi^2$,
one finds that the whole contribution $\delta_\phi \ln \chi = \ln
\chi_1 + \ln \chi_2$ consists of an exchange term $\ln \chi_1$ and a
mean-field-like term $\ln \chi_2$. The latter is given by
\begin{eqnarray}
    \ln \chi_2  =
    \frac{\phi^2 \T h^2}{2 \pi T_K^2} \frac{V}{\sinh(V/2T)}
    \left\{ [\cos(\mu) e^{-i\lambda} - 1] e^{V/2T} +
    [\cos(\mu) e^{i\lambda} - 1] e^{-V/2T} \right\} \, .
\end{eqnarray}
The mean-field-like contribution is mere combination of
spin-polarized terms. However, the exchange term contains new
pieces,
\begin{eqnarray}\label{lnchi1}
    \ln \chi_1 & = &
    \frac{\phi^2 \T V}{24 \pi T_K^2} \frac{V^2 + 4 (\pi
    T)^2}{\sinh(V/2T)} \left\{ [\cos(\mu) e^{-i\lambda} - 1] e^{V/2T} +
    [\cos(\mu) e^{i\lambda} - 1] e^{-V/2T} \right\} \nonumber \\
& + &
    \frac{\phi^2 \T V}{12 \pi T_K^2} \frac{V^2 + (\pi
    T)^2}{\sinh(V/T)} \left\{ [e^{-2i\lambda} - 1] e^{V/T} +
    [e^{2i\lambda} - 1] e^{-V/T} \right\}
+
    \frac{\phi^2 \T \pi T^3}{6 T_K^2} \left[ \cos(2\mu) - 1\right]\,
    .
\end{eqnarray}
The first contribution describes single electron tunneling and has
the same structure as the terms resulting from the scattering
Hamiltonian. The second part can be interpreted as electron-pair
tunneling. The fact that this term does not contain the spin
counting field is not surprising. It accomodates doubled counting
field which according to [\onlinecite{gogolin06}] can be ascribed
to electron pair tunneling. It is natural to expect that the spins
of this pair are oriented opposite to each other, so that the spin
current fluctuations are suppressed. The last term of
(\ref{lnchi1}) is a thermal, voltage-independent contribution
describing the transfer of two spins. Finally, the cross-term,
which is proportional to $\alpha \phi$, is given by
\begin{eqnarray}
    \delta_{\alpha\phi} \ln \chi  =
    \frac{\alpha\phi \T h^2}{\pi T_K^2} \frac{V}{\sinh(V/2T)}
    \left\{ [\cos(\mu) e^{-i\lambda} - 1] e^{V/2T} +
    [\cos(\mu) e^{i\lambda} - 1] e^{-V/2T} \right\}\, .
\end{eqnarray}
\end{widetext}

\section{Comparison of the results}
So far, we have analyzed the spin transport statistics for two
distinct realizations of the Kondo system: (i) the Toulouse point
corresponds to a certain quite special choice of the coupling
constants $J_i$ and (ii) the strong-coupling regime effectively
describing the system in the case $J \rightarrow \infty$, which,
according to the RG analysis is reached at low energies. The purpose
of this Section is to establish a connection between these two
situations.

There are several clear differences between the two regimes which
lead to different transport statistics. First, the strong-coupling
fixed point still possesses two symmetries which are broken in the
Toulouse fixed point by setting $J^{LR}_z = J^{RL}_z = 0$ and
$\Gamma_- \neq 0$. Moreover, in the above calculations at the
Toulouse point, the magnetic field was only applied to the dot. This
scheme cannot be easily implemented in the unitary limit since the
dot level, being perfectly screened, effectively disappears from the
system. Thus the magnetic field dependence in Section
\ref{unitarylimit} is referring to lead electrons. For this reason
we shall only compare the statistics at zero magnetic field.

From the Toulouse point calculation for $\Delta=\Gamma_-=0$, we
obtain the following expression for the combined statistics up to
order $V^3$,
\begin{equation}\label{Toulouse1}
    \ln \chi(\mu,\lambda) =
    2 i N \lambda + \frac{\T V^3}{6 \pi \Gamma_\perp^2}
    \left(e^{-2i\lambda} - 1\right) \, .
\end{equation}
The corresponding strong-coupling result for $h=0$ reads
\begin{eqnarray}\label{Kondo1}
    \ln \chi(\mu,\lambda)
& = &
    2 i N \lambda + \frac{(\alpha^2 + \phi^2) \T V^3}{12\pi T_K^2}
    \left[ \cos(\mu) e^{-i\lambda} - 1\right] \nonumber \\
& + & \frac{\phi^2 \T V^3}{6\pi T_K^2}
    \left(e^{-2i\lambda} - 1\right) \, .
\end{eqnarray}
Naturally, in the linear order in $V$ these two terms are
equivalent. Differences arise in the next-to-leading terms, most
notable is the $\mu$-dependence of (\ref{Kondo1}). The physical
reason for this effect is the following: by discarding the
$J^{LR}_z$ and $J^{RL}_z$ coupling terms at the Toulouse point,
spin-flip tunneling is the only transport process available in the
Toulouse limit. Under these conditions it is energetically favorable
to tunnel electron pairs with opposite spins since this process
effectively leaves the Kondo singlet state untouched. On the
contrary, in the true unitary limit where $J^{LR/RL}_z$ are finite,
electron tunneling without the impurity spin-flip is allowed, thus
leading to a contribution containing a single $\lambda$ field which
is also explicitly $\mu$-dependent. Another insight which is worth
mentioning is the Kondo temperature estimate $T_K \sim
\Gamma_\perp$, which coincides with the original proposal of
[\onlinecite{schiller98}].

\section{Conclusion}

To summarize, we have investigated the combined statistics of charge
and spin currents through a Kondo dot at the Toulouse point as well
as in the unitary limit. In the Toulouse limit at zero temperature,
the spin current noise grows linearly with $V$ while the spin
current is subleading. The spin transfer statistics is trinomial in
the linear response regime. At finite temperatures, we have
established the Johnson-Nyquist formula with an effective
transmission coefficient for this type of transport. We have
calculated explicitly the combined charge and spin current
correlations, which turn out to change sign depending on system
parameters. Furthermore, the spin-resolved charge transfer
statistics is shown to be binomial, independent on the spin
orientation and governed by a single effective transmission
coefficient.

In the strong-coupling limit the spin-resolved FCS has helped to
reveal a very rich structure of the transport processes. In addition
to the expected spin-isotropic single electron tunneling, the
contribution of electron pairs tunneling in singlet spin
configuration can be clearly made out. Moreover, the FCS contains a
part entirely due to spin currents which surprisingly does not
depend on the applied bias voltage and which vanishes at zero
temperature.

 TLS and AK
are supported by DAAD and DFG under grant No. KO 2235/2. AK is
also supported by the Feodor Lynen program of the Alexander von
Humboldt foundation.

\bibliography{bibliography}

\end{document}